\documentclass[format=sigconf,anonymous=false,review=false,screen=true]{acmart}

\AtBeginDocument{%
  \providecommand\BibTeX{{%
    \normalfont B\kern-0.5em{\scshape i\kern-0.25em b}\kern-0.8em\TeX}}}

\copyrightyear{2023}
\acmYear{2023}
\setcopyright{acmlicensed}\acmConference[SIGIR '23]{Proceedings of the 46th International ACM SIGIR Conference on Research and Development in Information Retrieval}{July 23--27, 2023}{Taipei, Taiwan}
\acmBooktitle{Proceedings of the 46th International ACM SIGIR Conference on Research and Development in Information Retrieval (SIGIR '23), July 23--27, 2023, Taipei, Taiwan}
\acmPrice{15.00}
\acmDOI{10.1145/3539618.3591909}
\acmISBN{978-1-4503-9408-6/23/07}

\makeatletter
\def\blfootnote{\gdef\@thefnmark{}\@footnotetext}
\makeatother

\usepackage{graphicx}
\usepackage[caption=false]{subfig}
\usepackage{color}
\usepackage{algorithm}
\usepackage{algorithmic}

\usepackage{microtype}
\usepackage{amsmath,amsfonts}
\usepackage{mathtools}
\usepackage{mathrsfs}
\usepackage{bm}
\usepackage{booktabs, multirow}
\usepackage{makecell}
\usepackage{stfloats}
\usepackage{xcolor}
\usepackage{arydshln}
\usepackage{pifont}
\usepackage{tikz}

\usepackage{multirow}
\usepackage[capitalize,noabbrev]{cleveref}
\usepackage{amsthm}
\usepackage{balance}
\theoremstyle{definition}

\newcommand*{\circled}[1]{\lower.7ex\hbox{\tikz\draw (0pt, 0pt)%
    circle (.5em) node {\makebox[1em][c]{\small #1}};}}

\begin{document}

\title{FedAds: A Benchmark for Privacy-Preserving CVR Estimation with Vertical Federated Learning}

\author{Penghui Wei}
\affiliation{%
  \institution{Alibaba Group}
  \city{} 
  \country{} 
}
\email{wph242967@alibaba-inc.com}

\author{Hongjian Dou}
\affiliation{%
  \institution{Alibaba Group}
  \city{} 
  \country{} 
}
\email{hongjian.dhj@alibaba-inc.com}

\author{Shaoguo Liu$^{\scriptscriptstyle *}$}
\affiliation{%
  \institution{Alibaba Group}
  \city{} 
  \country{} 
}
\email{shaoguo.lsg@alibaba-inc.com}

\author{Rongjun Tang$^{ \dagger}$}
\affiliation{%
  \institution{The Chinese University of Hong Kong, Shenzhen}
  \city{} 
  \country{} 
}
\email{rongjuntang@link.cuhk.edu.cn}

\author{Li Liu}
\affiliation{%
  \institution{The Thrust of Artificial Intelligence, The Hong Kong University of Science and Technology
(Guangzhou)}
  \city{} 
  \country{} 
}
\email{avrillliu@hkust-gz.edu.cn}

\author{Liang Wang}
\affiliation{%
  \institution{Alibaba Group}
  \city{} 
  \country{} 
}
\email{liangbo.wl@alibaba-inc.com}

\author{Bo Zheng}
\affiliation{%
  \institution{Alibaba Group}
  \city{} 
  \country{} 
}
\email{bozheng@alibaba-inc.com}

\thanks{$^{\dagger}$Work done during internship at Alibaba Group. $^{\scriptscriptstyle *}$Correspondence to: S. Liu.}

\renewcommand{\authors}{Penghui Wei, Hongjian Dou, Shaoguo Liu, Rongjun Tang, Li Liu, Liang Wang and Bo Zheng}
\renewcommand{\shortauthors}{Penghui Wei et al.}

\begin{abstract}
Conversion rate (CVR) estimation aims to predict the probability of conversion event after a user has clicked an ad. Typically, online publisher has user browsing interests and click feedbacks, while demand-side advertising platform collects users' post-click behaviors such as dwell time and conversion decisions. To estimate CVR accurately and protect data privacy better, vertical federated learning (vFL) is a natural solution to combine two sides' advantages for training models, without exchanging raw data. Both CVR estimation and applied vFL algorithms have attracted increasing research attentions. However, standardized and systematical evaluations are missing: due to the lack of standardized datasets, existing studies adopt public datasets to \textit{simulate} a vFL setting via hand-crafted feature partition, which brings challenges to fair comparison. 
We introduce \textsf{FedAds}, the first benchmark for CVR estimation with vFL, to facilitate standardized and systematical evaluations for vFL algorithms. It contains a large-scale real world dataset collected from Alibaba's advertising platform, as well as systematical evaluations for both effectiveness and privacy aspects of various vFL algorithms. Besides, we also explore to incorporate unaligned data in vFL to improve effectiveness, and develop perturbation operations to protect privacy well. We hope that future research work in vFL and CVR estimation benefits from the \textsf{FedAds} benchmark. 

\end{abstract}

%
%

\ccsdesc[500]{Information systems~Online advertising}
\ccsdesc[500]{Computing methodologies~Neural networks}

\keywords{Ad Ranking, Vertical Federated Learning, Deep Generative Model}

\maketitle

\section{Introduction}\label{intro}
Nowadays, more and more Web applications employ machine learning models to provide personalized services for users to meet their preferences. 
Conversion rate (CVR) estimation is a foundational module in online recommendation and advertising systems, which aims to predict the probability of conversion event (e.g., purchase behavior in e-commerce ads) after a user has clicked an item/ad. It plays a crucial in candidate ranking and ad bidding strategies.

\begin{figure}[t]
\centering
\centerline{\includegraphics[width=\columnwidth]{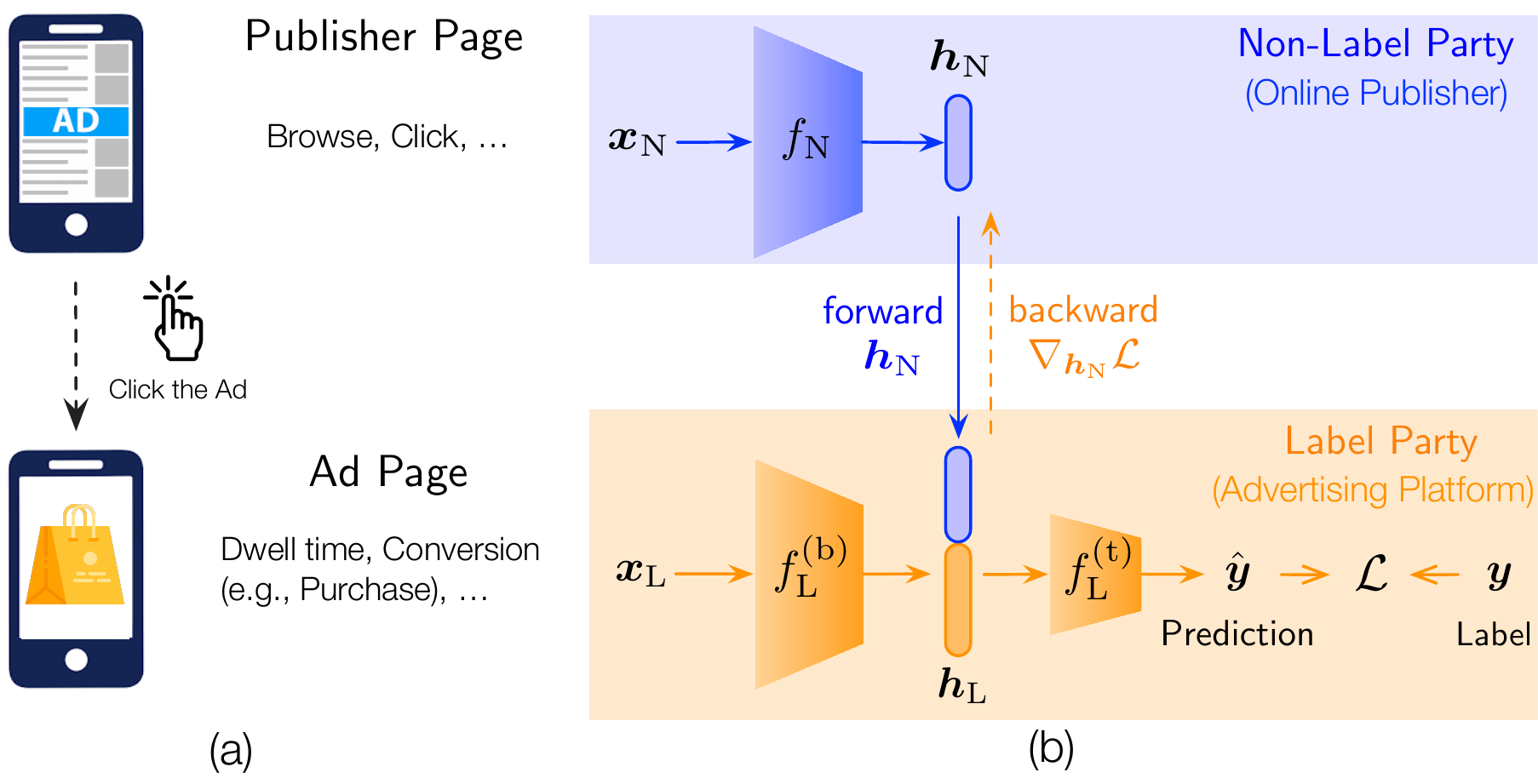}}
\vspace{-1em}
\caption{(a) The feedback behaviors after a user browses an ad. (b) Vertical federated learning framework for training conversion rate estimation model. Online publisher and  advertising platform provide the collected user feedback data and collaboratively train the model. }
\label{fig:splitlearning}
\end{figure}

Figure~\ref{fig:splitlearning} (a) gives an illustration of user behaviors in online advertising. If a user browses an ad at the {publisher page} and then clicks on it, the user will arrive at an {ad landing page}. Follow-up behaviors including conversion decisions may happen. Therefore, the \textit{online publisher} has user browsing interests and click feedbacks, and the demand-side \textit{advertising platform} collects users' post-click behaviors on ad pages such as dwell time and conversion. CVR estimation models are trained using collected data about user profiles, user behaviors and item/ad properties, with the goals of improving user experiences and increasing platform revenue. However due to the potential abuse and leak of user data, continued concerns regarding data privacy and security are raised. 

User behaviors collected by online publisher or advertising platform are private information and should be protected. To estimate CVR accurately and protect data privacy better, vertical federated learning (vFL)~\cite{vepakomma2018split,yang2019federated} is a natural solution for model training, which combines two sides' advantages without exchanging raw data. As a branch of federated learning (FL)~\cite{mcmahan2017communication} techniques, vFL can collaboratively train a model using feature-partitioned data from multiple participants via exchanging \textit{intermediate results} (e.g., hidden representations and gradients) rather than \textit{raw data} (e.g., features and labels). 
Figure~\ref{fig:splitlearning} (b) shows the vFL framework for training a neural network based CVR estimation model. There are two participants, where the \textbf{non-label party} is online publisher and the \textbf{label party} is advertising platform that owns conversion labels. The training data for vFL is \textbf{feature-partitioned}: before model training, the two participants first perform private set intersection (PSI)~\cite{kolesnikov2016efficient} to obtain an aligned sample ID set, and each sample's features come from both non-label party (e.g., behaviors on publisher page) and label party (e.g., behaviors on ad page). 

The whole model is split into two submodels owned by non-label and label parties respectively. During the \textit{forward} pass, for a given input sample, the non-label party's submodel sends a \textit{hidden representation} to the label party. Then the label party combines such representation with its own one, and produces the predicted conversion probability and cross-entropy loss according to the sample's conversion label. During the \textit{backward} propagation, the label party computes \textit{gradient} w.r.t. the non-label party's hidden representation and sends it back, thus the update of non-label party's submodel parameters depends on the label party.

Both CVR estimation~\cite{ma2018esmm,wei2021autoheri,guo2021enhanced,xu2022ukd} and applied vFL algorithms~\cite{castiglia2022SSVFL,kang2022fedcvt,li2022vflmpd,he2022FedHSSL,li2022jpl,li2022label,sun2022label,fu2022label} have attracted increasing research attentions. 
Specifically, to improve privacy-preserving CVR estimation, two research problems in vFL should be tackled. 
The first is about \textbf{effectiveness}: Traditional vFL  training procedure only employs \textit{aligned} samples among multiple participants, however the size of aligned samples is quite limited, which restricts the model performance. Various approaches based on self-supervised learning~\cite{castiglia2022SSVFL,he2022FedHSSL} and semi-supervised learning~\cite{kang2022fedcvt,li2022vflmpd,li2022jpl} are proposed to explore the potential usage of a large amount of unaligned samples owned by each party for improving vFL. 
The second is about \textbf{privacy}: Although vFL only exchange intermediate results rather than raw features and labels, recent studies revealed that it still suffers from privacy leakage risks such as label inference attack, which means that a honest-but-curious non-label party can successfully infer private labels. To defend against such attack, many studies focus on random perturbation based approaches~\cite{li2022label,sun2022label,fu2022label} for protecting private label information. 

However, standardized datasets and systematical evaluations for vFL algorithms are missing: due to the lack of standardized datasets, existing studies usually adopt public datasets to \textit{simulate} a vFL experiment setting via hand-crafted feature partition, rather than adopting datasets from real-world vFL applications. This situation brings challenges to fair comparison of various models and hinders further research of vFL and privacy-preserving CVR estimation. As a result, above-mentioned vFL algorithms are not compared under the same dataset and evaluation procedure currently. Therefore, there is a great need for a comprehensive benchmark to facilitate vFL research. 

In this paper we introduce \textsf{FedAds}, the first benchmark for privacy-preserving CVR estimation with vFL, to facilitate systematical evaluations for vFL algorithms. It contains 1) a large-scale real-world dataset from our online advertising platform, collected from an ad delivery business relying on vFL-based ranking models, as well as 2) systematical evaluations for both effectiveness and privacy aspects of various neural network based vFL algorithms through extensive experiments. Besides, to improve vFL effectiveness, we explore to incorporate unaligned data via generating unaligned samples' feature representations using generative models. To protect privacy well, we also develop perturbation based on mixup and projection operations. We hope that future research work in both vFL algorithms and CVR estimation benefits from our \textsf{FedAds} benchmark.

The main contributions of this work are:
\begin{itemize}
    \item We provide a real-world CVR estimation dataset collected from our ad delivery business relying on vFL-based ranking models. To our knowledge, this is the first large-scale dataset for vFL research.  
    \item We conduct systematical evaluations to recently proposed vFL algorithms for effectiveness and privacy aspects respectively on the proposed dataset, which promotes fair comparison of various studies. 
    \item  We propose two approaches for incorporating unaligned samples and protecting private label information in vFL respectively, and experiments on the proposed dataset verify their performance. 
\end{itemize}

\section{Preliminaries}\label{pres}

\subsection{Conversion Rate Estimation}
The goal of a post-click conversion rate (CVR) estimation model $f(\cdot)$ is to produce the conversion probability if a user has clicked an ad: $\hat p_\mathrm{CVR}=f(\boldsymbol x)$, where $\boldsymbol x$ denotes input feature of a sample. The model $f(\cdot)$ is trained using click log $\mathcal D=\left\{\left(\boldsymbol{x}, y\right)\right\}$ with cross-entropy loss, where $y\in \{0,1\}$ is the sample's binary conversion label. 
CVR estimation models are usually deployed to the ranking module of online advertising and recommendation systems, and they are crucial for improving user experiences, satisfying advertiser demands and increasing the revenue of advertising platform.

\subsection{Vertical Federated Learning}\label{SL}
We first give a brief introduction to two-party vFL framework, and then discuss the issues in traditional vFL. 

\subsubsection{\textbf{Two-Party vFL}}\label{sec:vfl}
We consider the overall framework in~\cref{fig:splitlearning} (b). Without loss of generality, we assume that there are two separate participants, namely non-label party and label party. They cooperate with each other to learn a model $f:\mathcal X\rightarrow \mathcal Y$. 
The feature space
$\mathcal X=\mathcal X_\mathrm{N}\cup \mathcal X_\mathrm{L}$ is composed of two parts, where $\mathcal X_\mathrm{N}$ / $\mathcal X_\mathrm{L}$ represents the feature subspace owned by \underline{N}on-label party / \underline{L}abel party. 
The label party owns the one-hot label space $\mathcal{Y}$. 

The whole model $f(\cdot)$ is split into two submodels $f_\mathrm{N}(\cdot)$ and $f_\mathrm{L}(\cdot)$ which are owned by non-label party and label-party respectively: 
\begin{equation}
    f \coloneqq f_\mathrm{N} \circ f_\mathrm{L}\,,
\end{equation}
here $f_\mathrm{N}(\cdot)$ is the non-label party's submodel which produces hidden representation for each sample and sends it to label party, 
while $f_\mathrm{L}(\cdot)$ is the label party's submodel that produces the predicted probability distribution.  

Before model training, the two parties first perform PSI to obtain an \textbf{aligned} sample set $\mathcal D_\mathrm{aligned}=\{(\boldsymbol x_\mathrm{N}, \boldsymbol x_\mathrm{L}, \boldsymbol{y})\}$, 
where $\boldsymbol x_\mathrm{N}$ and $\boldsymbol x_\mathrm{L}$ denote the sample's feature provided by two parties respectively, and its label $\boldsymbol y$ is a one-hot vector. 
Consider the forward pass, for each sample the model $f(\cdot)$ outputs the predicted distribution $\hat {\boldsymbol y}$, and then computes the cross-entropy loss $\mathcal L$: 
\begin{equation}
    \boldsymbol h_\mathrm{N} = f_\mathrm{N}(\boldsymbol x_\mathrm{N})\,,\quad \boldsymbol h_\mathrm{L} = f_\mathrm{L}^{(\mathrm{b})}(\boldsymbol x_\mathrm{L})
\end{equation}
\begin{equation}
    \boldsymbol l  = f_\mathrm{L}^{(\mathrm{t})} \bigl(   \boldsymbol h_\mathrm{N}, \boldsymbol{h}_\mathrm{L} \bigr) \,, \quad 
    \hat {\boldsymbol y} = \mathrm{softmax}(\boldsymbol l)  
\end{equation}
\begin{equation}\label{eq:vfl_loss} 
    \mathcal{L} = -\boldsymbol y^{\top}\log \hat {\boldsymbol y} 
\end{equation}
where $\boldsymbol h_\mathrm{N}$ is the hidden representation that the non-label party sends to the label party, and the output layer of $f_\mathrm{N}(\cdot)$ is also known as \textbf{cut layer}. The label party's submodel $f_\mathrm{L}(\cdot)$ is composed of a bottom part $f_\mathrm{L}^{(\mathrm{b})}(\cdot)$ and a top part $f_\mathrm{L}^{(\mathrm{t})}(\cdot)$. 

For the backward pass, the label party's submodel $f_\mathrm{L}(\cdot)$ is updated normally based on the gradient of loss $\mathcal L$ w.r.t. the submodel parameters.  
To update the parameters of non-label party's submodel $f_\mathrm{N}(\cdot)$, the label party further computes the gradient w.r.t. the hidden representation $\boldsymbol{h}_\mathrm{N}$, and sends it back to the non-label party:
\begin{equation}\label{eq:gradient}
    \boldsymbol{g} \coloneqq \nabla_{{\boldsymbol{h}_\mathrm{N}}} \mathcal{L} = \frac{\partial \boldsymbol{l}}{\partial \boldsymbol{h}_\mathrm{N}} \frac{\partial \mathcal L}{\partial \boldsymbol{l}} = \frac{\partial \boldsymbol{l}}{\partial \boldsymbol{h}_\mathrm{N}}(\hat {\boldsymbol y} - {\boldsymbol y})\,.
\end{equation}
After receiving the gradient $\boldsymbol{g}$, the non-label party computes the gradients of the remaining submodel parameters using chain rule, and thus continues the backward pass to update the submodel $f_\mathrm{N}(\cdot)$.

In the above training process, the two participants do not share their raw data (features $\boldsymbol x_\mathrm{N}$, $\boldsymbol x_\mathrm{L}$ and label $\boldsymbol{y}$) to each other. On the contrary, they only exchange intermediate results $\boldsymbol{h}_\mathrm{N}$ and $\boldsymbol{g}$. 

\subsubsection{\textbf{Issues of Effectiveness and Privacy}}\label{sec:vfl_issues}
vFL has been successfully applied in healthcare informatics~\cite{vepakomma2018split}, computational advertising~\cite{li2022label} and many other domains. To further improve vFL algorithms, recent studies pay more attention to the following perspectives: effectiveness and privacy. 

\vspace{0.5em}
\textbf{Effectiveness: Limited Aligned Samples.}

The training procedure of traditional vFL algorithms relies on aligned, feature-partitioned data. That is, each participant first provides its own sample ID set, and then a PSI process is adopt to align samples from all participants, and finally produces the \textit{intersection} sample set, namely aligned samples. In the aligned data, for a specific sample, each participant owns a part of feature of it, and all participants need to collaboratively train the vFL model. 

We can see that the vFL model performance greatly depends on the size of aligned samples. However the size is usually limited, which restricts the effectiveness of vFL models. 
To tackle this, a direction is to make use of local samples of each participant, namely unaligned samples. 
In the case of CVR estimation, the advertising platform usually also collects conversion events from in-station ads (which are not displayed on the extra publishers). These local data can be used as auxiliary training samples to enhance the model trained on aligned samples only. To exploit such unaligned samples that only have partly features into the vFL training framework, we focus on how to synthesize the features of other participants.

\vspace{0.5em}
\textbf{Privacy: Potential Label Leakage.}

vFL is often considered to be privacy-oriented, because during training process participants only exchange intermediate hidden representations and gradients rather than raw features and labels. 
However, recent studies revealed that it still suffers from potential privacy leakage risks: (1) label inference attack~\cite{li2022label,sun2022label,fu2022label}, which means that a honest-but-curious non-label party can successfully infer private labels, and (2) input reconstruction attack~\cite{vepakomma2018reducing,sun2021defending}, which means that the label party can reconstruct the raw input features of the non-label party. 

In this work we focus on defending against label inference attacks, aiming to guarantee that the private label information owned by label party can be protected. Specifically, although the label party only sends gradients to the non-label party during training, from Equation~\ref{eq:gradient} we can observe that the mathematical expression of the gradient $\boldsymbol{g}$ w.r.t. hidden representation contains label information $\boldsymbol{y}$, which results in potential label leakage. 

During the training procedure of vFL models, privacy-preserving computing techniques can be applied to protect private information. For instance, the open-source framework \textsf{EFLS}\footnote{\url{https://github.com/alibaba/Elastic-Federated-Learning-Solution}} from Alibaba Group provides APIs of cryptography-based homomorphic encryption~\cite{zhang2020additively} and perturbation-based  differential privacy~\cite{abadi2016deep}. In this work we focus on random perturbation-based algorithms.

\section{Proposed Benchmark}
Currently, the lack of a comprehensive benchmark brings challenges to fair comparison of various algorithms, and also hinders further research for tacking the effectiveness and privacy issues of vFL. To address this, we introduce \textsf{FedAds}, the first benchmark for privacy-preserving CVR estimation with vFL, to facilitate systematical evaluations for vFL algorithms. 

\subsection{Overview}
The \textsf{FedAds} benchmark contains:
\begin{itemize}
    \item A large-scale dataset from Alibaba's advertising platform, collected from the log of an ad delivery business relying on vFL-based ranking models. Details  in Section~\ref{sec:dataset}. 
    \item Systematical evaluations for both effectiveness and privacy aspects of various vFL algorithms. Details in Section~\ref{sec:exp}. 
\end{itemize}

We release the \textsf{FedAds} benchmark at the page \url{https://github.com/alibaba/Elastic-Federated-Learning-Solution/tree/FedAds}. 

\subsection{Real-World Dataset Construction}\label{sec:dataset}
We first introduce the background of the collected data, and then give statistics and features of the dataset. 
\subsubsection{\textbf{Data Description}} 
\begin{figure}[t]
\centering
\centerline{\includegraphics[width=0.9\columnwidth]{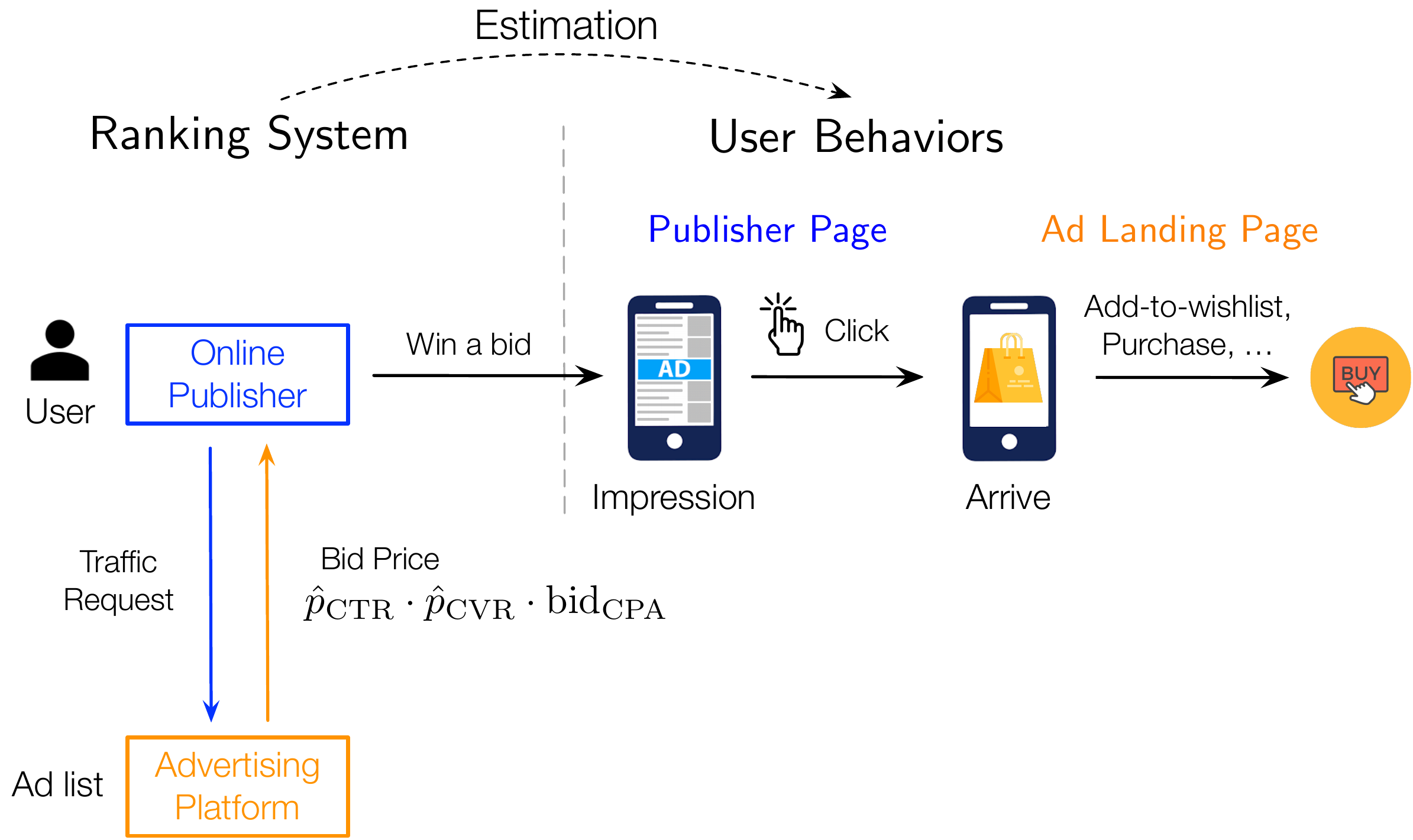}}
\caption{Brief illustration of the ranking stage in ad delivery procedure and user behaviors. The proposed dataset is built upon the click log collected from this procedure. }
\vspace{-0.5em}
\label{fig:collect}
\end{figure}

The dataset is built upon the click log of our e-commerce ad delivery business, in which both the online publisher and the advertising platform belong to Alibaba Group. Although the two parties belong to the same company, they still cannot share user behavior information to each other. 
Specifically, the online publisher is a mobile app that contains ad positions. As shown in Figure~\ref{fig:collect}, the advertising platform bids for ad impressions in real-time, and for each request the predicted CVR score $\hat p_\mathrm{CVR}$ is a key factor in the bid price. 
If an ad from the advertising platform wins a bid, it will be displayed to the user. The user will arrive at another e-commerce mobile app that manages the ad landing page if he/she clicks on the ad, and may take further behaviors such as add-to-wishlist and purchase. 

The above ad delivery business is a typical application of vFL, where the online publisher and the advertising platform collaboratively train the CVR estimation model for ranking candidate ads and improving the delivery performance.  
Therefore, we believe that further vFL research can benefit from our benchmark.

\begin{table}[b]
\caption{Statistics of the dataset.}
\centering
\begin{tabular}{ccccc}
\toprule
{Split}  & {\# Samples}  & {\# Users} & {\# Ads}  & {CVR} \\
\midrule
Training+Test   & 11.3 mil. & 4.1 mil. & 1.3 mil. & 0.6\%\\
Training        & 10.0 mil. & 3.7 mil. & 1.2 mil. & 0.6\%\\
Test            & 1.3  mil. & 0.7 mil. & 0.4 mil. & 0.6\% \\
\bottomrule
\end{tabular}
\label{tab:stat}
\end{table}

\subsubsection{\textbf{Dataset Construction}}
We built the dataset based on the above collected data. 
Specifically, we collect 1-month consecutive user click events of the delivery business, and each sample in the dataset is corresponding to a unique click event. We record context information for each sample, such as the timestamps of  request and click event. Generally, the dataset is composed of features from both parties, and conversion labels from label party.

\vspace{0.5em}
\textbf{Conversion label.}

A sample's label is set to 1 if the user purchases the item described by the clicked ad, where the attribution window is set to 24 hours. Here we employ last-touch attribution, which means that if a user clicks on the ad multiple times and finally purchases the item, we regard that this conversion event is attributed by the last click event.

\vspace{0.5em}
\textbf{Features and processing.}

The feature set for each sample consists of two parts: one part from the label party (i.e., advertising platform) and another one from the non-label party (i.e., online publisher). Specifically, 
\begin{itemize}
    \item Features from label party: we construct user-side, ad-side and context features. 
    \begin{itemize}
            \item [-] The user-side features contain user profile information (e.g., user ID and activity level), as well as purchase-related behaviors such as user's historial purchased items. 
            \item [-] The ad-side features contain profile information (e.g., ad ID, item ID, item's brand ID and price level), as well as ad statistics like historial click/conversion count level. 
            \item [-] Context feature is the timestamp of conversion event.
    \end{itemize}
    \item Features from non-label party: similarly, we construct user-side, ad-side and context features. 
    \begin{itemize}
            \item [-] The user-side features are  click-related behaviors such as user's historial clicked ads. 
            \item [-] The ad-side features are statistical information such as historial impression count level. 
            \item [-] Context feature is the timestamp of click event. 
    \end{itemize}
\end{itemize}

In summary, there are 16 features owned by label party and 7 features owned by non-label party. 
For the considerations of fair comparison and removing personal identifiable information, in our dataset we release the processed features rather than original values. Specifically, for discrete features we map the original values to IDs. For each continuous feature, we perform equi-frequency discretization to transform the original values to bin IDs.

\subsubsection{\textbf{Statistics}}
Table~\ref{tab:stat} lists the statistics of our constructed dataset. 
Totally the dataset contains 11.3 million samples, and to our knowledge it is the largest public dataset for evaluating CVR estimation models and vFL algorithms. 
We split it to training set and test set based on click timestamp, where the last week's samples are selected for test set. Details about how to use the dataset for evaluating effectiveness and privacy are introduced in Section~\ref{sec:exp}.

We compare our proposed dataset with current commonly-used datasets in vFL research in Table~\ref{tab:vfldatasets}. 
More importantly, our dataset is constructed from real world applications, thus we do not need to simulate a vFL experiment setting.

\begin{table}[b]
\caption{Comparison of datasets.}
\centering
\begin{tabular}{ccc}
\toprule
{Split}  & {\# Samples} & {Type} \\
\midrule
BHI~\cite{bhi}                  & 0.3 mil. & Image \\
Yahoo Answers~\cite{yahoo}        & 1.5 mil. & Text \\

Give Me Some Credit~\cite{givemesomecredit}  & 0.3 mil. & Tabular \\
Epsilon~\cite{epsilon}              & 0.5 mil. & Tabular \\

Avazu~\cite{avazu}                & 4 mil. & Tabular\\ 

Criteo~\cite{criteo}               & 4.5 mil. & Tabular \\
\textsf{FedAds} (Ours)  & 11.3 mil. & Tabular\\
\bottomrule
\end{tabular}
\label{tab:vfldatasets}
\end{table}

\begin{figure}[t]
\centering
\centerline{\includegraphics[width=\columnwidth]{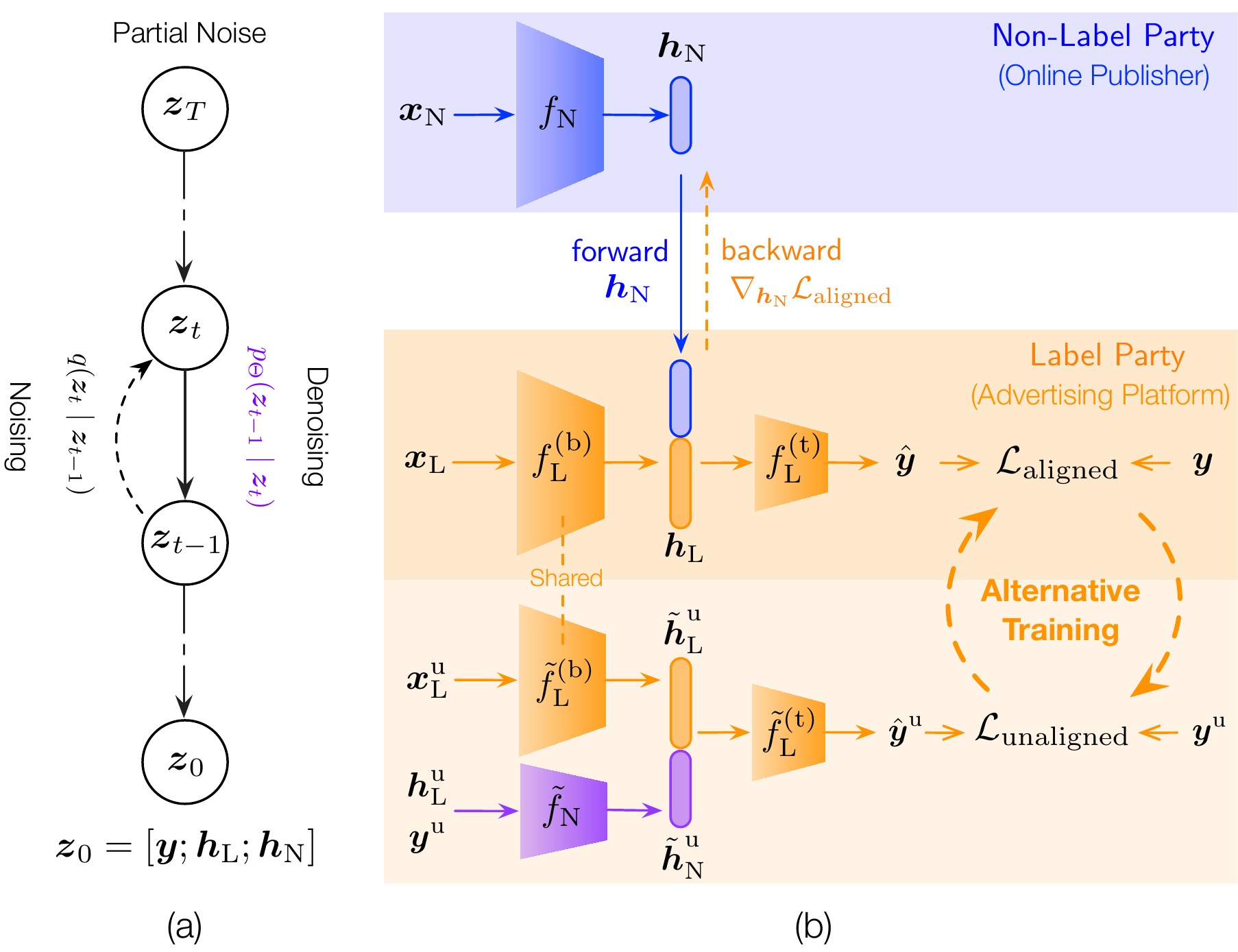}}
\caption{(a) Learning a conditional diffusion model for generating federated embeddings of label party's  unaligned samples. (b) The \textsf{Diffu-AT} framework for exploiting unaligned samples in vFL training. }
\label{fig:diffmodel}
\end{figure}

\section{Methodology}
Before the systematical evaluations of existing vFL models, we propose two approaches for improving effectiveness and privacy. 

\subsection{Exploiting Label Party's Unaligned Samples}
As statement in Section~\ref{sec:vfl_issues}, the limited size of aligned samples $\mathcal D_\mathrm{aligned}=\{(\boldsymbol x_\mathrm{N}, \boldsymbol x_\mathrm{L}, \boldsymbol{y})\}$ restricts the effectiveness of vFL models. From the perspective of advertising platform, a nature way to tackle this problem is to incorporate its local and unaligned samples $\mathcal D_\mathrm{unaligned} = \left\{ \left( \boldsymbol{x}_\mathrm{L}^{\mathrm{u}}, \boldsymbol{y}^{\mathrm{u}}  \right) \right\}$ into the vFL training procedure. 

However, the challenge of exploiting label party's unaligned samples in vFL is that they do not have non-label party's features (i.e., $\boldsymbol{x}_\mathrm{N}^{\mathrm{u}}$ is missing). We propose \textsf{Diffu-AT}, an enhanced vFL training framework which first generates missing feature with a diffusion model, and then performs alternatively training to incorporate unaligned samples into the traditional vFL framework.

\subsubsection{\textbf{Generating Federated Embedding $\tilde {\boldsymbol{h}}_\mathrm{L}^{\mathrm{u}}$ with Conditional Diffusion Model}}
As shown in~\cref{fig:splitlearning}, the key of traditional vFL training procedure is that the non-label party sends the representation $\boldsymbol{h}_\mathrm{N}$ (named \textbf{federated embedding} in this work) to enhance the prediction. 
To effectively incorporate unaligned samples into vFL, we employ deep generative models to synthesize federated embeddings $\tilde {\boldsymbol{h}}_\mathrm{N}^{\mathrm{u}}$ for those samples.

\vspace{0.5em}
\textbf{Problem formulation.}

We formulate the synthesis process is to learn a generation model $\tilde{f}_\mathrm{N}(\cdot)$, which can generate a federated embedding $\tilde {\boldsymbol{h}}_\mathrm{N}^{\mathrm{u}}$ given label party's feature $\boldsymbol{x}_\mathrm{L}^{\mathrm{u}}$ and label $\boldsymbol{y}^{\mathrm{u}}$, namely $ \tilde {\boldsymbol{h}}_\mathrm{N}^{\mathrm{u}} = \tilde{f}_\mathrm{N} \left(\boldsymbol{x}_\mathrm{L}^{\mathrm{u}}, \boldsymbol{y}^{\mathrm{u}}   \right)$.

\vspace{0.5em}
\textbf{Step 1: vFL pretraining.}

To this end, we first perform vFL training, that is, pretrain a vFL model $f(\cdot)$ using aligned samples $\mathcal D_\mathrm{aligned}$ with the loss in Eq.~\ref{eq:vfl_loss}. 
Therefore, based on the pretrained vFL model, for each aligned sample we obtain its federated embedding ${\boldsymbol{h}}_\mathrm{N}$ and label party's representation ${\boldsymbol{h}}_\mathrm{L}$. 
Similarly, for each unaligned sample, we obtain its label party's representation ${\boldsymbol{h}}_\mathrm{L}^{\mathrm{u}}$.

Next, we use the data $\left\{ ({\boldsymbol{h}}_\mathrm{N}, {\boldsymbol{h}}_\mathrm{L}, \boldsymbol{y}) \right\}$ of aligned samples to learn the generation model $\tilde{f}_\mathrm{N}(\cdot)$, so as to perform inference on the data $\left\{\left(\boldsymbol{h}_\mathrm{L}^{\mathrm{u}}, {\boldsymbol{y}}^{\mathrm{u}}   \right)\right\}$ of unaligned samples to generate $\left\{ \tilde {\boldsymbol{h}}_\mathrm{N}^{\mathrm{u}} \right\}$.\footnote{Note that compared to previous problem formulation, we simplify the synthesis process via replacing the input $\boldsymbol{x}_\mathrm{L}^{\mathrm{u}}$ with $\boldsymbol{h}_\mathrm{L}^{\mathrm{u}}$.}

\vspace{0.5em}
\textbf{Step 2: Learning a conditional diffusion model as $\tilde{f}_\mathrm{N}(\cdot)$.}

Inspired by recent studies on diffusion-based deep generative models~\cite{sohl2015deep,ho2020denoising,lidiffusion,gongsequence}, we propose to synthesize unaligned samples' federated embeddings via a conditional diffusion model. 

Formally, we regard the concatenation of label party's representation $\boldsymbol{h}_\mathrm{L}$ and label ${\boldsymbol{y}}$ as the condition $\left[ {\boldsymbol{y}} ; \boldsymbol{h}_\mathrm{L} \right]$. 
We define the forward (noising) process with $T$ steps as: 
\begin{equation}
    \begin{aligned}
        q\left( \boldsymbol{h}_{\mathrm{N}, t} \mid \boldsymbol{h}_{\mathrm{N}, t-1} \right) &= \mathcal N\left( \boldsymbol{h}_{\mathrm{N}, t} ; \sqrt{1-\beta_t}\boldsymbol{h}_{\mathrm{N}, t-1}, \beta_t\mathbf{I}  \right) \\
        \boldsymbol{z}_t &\coloneqq \left[ {\boldsymbol{y}} ; \boldsymbol{h}_\mathrm{L} ; \boldsymbol{h}_{\mathrm{N}, t}  \right]
    \end{aligned}
\end{equation}
where $\boldsymbol{h}_{\mathrm{N},0} \coloneqq \boldsymbol{h}_{\mathrm{N}}$ and $\boldsymbol{z}_0 \coloneqq \left[ {\boldsymbol{y}} ; \boldsymbol{h}_\mathrm{L} ; \boldsymbol{h}_\mathrm{N}  \right]$, which means that at each step we incrementally add Gaussian noise on the part of $\boldsymbol{h}_\mathrm{N}$ only, while keeping the condition part unchanged. $\beta_t$ is a hyperparameter that controls the degree of noise at each step, and after $T$ steps the $\boldsymbol{h}_{\mathrm{N}, T}$ is approximately Gaussian. 

Further, the reverse (denoising) process is to reconstruct the original $\boldsymbol{z}_{0}$ given $\boldsymbol{z}_{T}$:
\begin{equation}
    p_\Theta(\boldsymbol{z}_{t-1} \mid \boldsymbol{z}_{t}) = \mathcal{N}\left( \boldsymbol{z}_{t-1} ; \boldsymbol{\mu}_{\Theta}(\boldsymbol{z}_{t}, t) , \boldsymbol{\Sigma}_{\Theta}(\boldsymbol{z}_{t}, t)  \right)
\end{equation}
where $\boldsymbol{\mu}_{\Theta}(\boldsymbol{z}_{t}, t)$ and $\boldsymbol{\Sigma}_{\Theta}(\boldsymbol{z}_{t}, t)$ are the predicted mean and standard deviation of $p_\Theta(\boldsymbol{z}_{t-1} \mid \boldsymbol{z}_{t})$ respectively, parameterized by learnable $\Theta$. 
The  objective is to maximize the marginal likelihood $\mathbb E_{q(\boldsymbol{z}_0)}\left[ \log p_{\Theta}(\boldsymbol{z}_0) \right]$, and it can be optimized using the variational lower bound~\cite{sohl2015deep}. 
In practice, we follow \textsf{DDPM}~\cite{ho2020denoising} to use a simplified form that the training stability has been empirically proved:
\begin{equation}
    \mathcal{L}_\mathrm{DDPM} = \sum_{t=1}^T \mathbb E_{q(\boldsymbol{z}_t \mid \boldsymbol{z}_0 )} \left[ \lVert  \tilde{\boldsymbol{\mu}}_t(\boldsymbol{z}_t,\boldsymbol{z}_0)  -  \boldsymbol{\mu}_{\Theta}(\boldsymbol{z}_{t}, t) \rVert^2 \right]
\end{equation}
where $\tilde{\boldsymbol{\mu}}_t(\boldsymbol{z}_t,\boldsymbol{z}_0)=\frac{\sqrt{\alpha_t}(1 - \bar{\alpha}_{t-1})}{1 - \bar{\alpha}_t} \boldsymbol{z}_t + \frac{\sqrt{\bar{\alpha}_{t-1}}\beta_t}{1 - \bar{\alpha}_t} \boldsymbol{z}_0$ is the mean of posterior $q(\boldsymbol{z}_{t-1} \mid \boldsymbol{z}_{t}, \boldsymbol{z}_{0})$, and here $\alpha_t=1-\beta_t, \bar{\alpha}_t=\prod_{i=1}^t\alpha_i$. 

After learning the diffusion model, we perform conditional generation to obtain each unaligned sample's federated embedding $\tilde {\boldsymbol{h}}_\mathrm{N}^{\mathrm{u}}$ given the conditional input $\boldsymbol{z}_T = \left[ {\boldsymbol{y}}^{\mathrm{u}} ; \boldsymbol{h}_\mathrm{L}^{\mathrm{u}}; \boldsymbol{h}_T   \right]$, where $ \boldsymbol{h}_T \sim \mathcal N(\boldsymbol{0}, \mathbf{I})$ is the initiate state. The generation repeats $T$ steps of denoising operation using the learned model $\Theta$ for sampling $\boldsymbol{z}_{T-1}, \boldsymbol{z}_{T-2},\ldots ,\boldsymbol{z}_{0}$ using $\boldsymbol{z}_{t-1}\sim p_\Theta(\boldsymbol{z}_{t-1} \mid \boldsymbol{z}_{t})$. Note that at each step, we replace the condition part of the generated $\boldsymbol{z}_t$ to the original condition $\left[ {\boldsymbol{y}}^{\mathrm{u}} ; \boldsymbol{h}_\mathrm{L}^{\mathrm{u}}\right]$. 
Finally at the last step we obtain $\tilde {\boldsymbol{h}}_\mathrm{N}^{\mathrm{u}}\coloneqq \boldsymbol{h}_0$ as the synthesized federated embedding, where $\boldsymbol{h}_0$ is the part in $\boldsymbol{z}_0$.

\subsubsection{\textbf{Alternative Training Framework}}\label{sec:at}
Now we have aligned samples $\mathcal D_\mathrm{aligned}=\{(\boldsymbol x_\mathrm{N}, \boldsymbol x_\mathrm{L}, \boldsymbol{y})\}$ as well as unaligned samples with synthesized federated embeddings $\tilde{\mathcal D}_\mathrm{unaligned} = \left\{ \left( \tilde{\boldsymbol{h}}_\mathrm{N}^{\mathrm{u}}, \boldsymbol{x}_\mathrm{L}^{\mathrm{u}}, \boldsymbol{y}^{\mathrm{u}}  \right) \right\}$.  

To effectively fuse them for learning an enhanced federated model that improves effectiveness, we propose to combine the two sample set in an alternative training fashion. As shown in Figure~\ref{fig:diffmodel} (b), the label party augments an auxiliary submodel $\tilde{f}_\mathrm{L}(\cdot)$ composed of $\left(\tilde{f}_\mathrm{L}^{(\mathrm{b})},  \tilde{f}_\mathrm{L}^{(\mathrm{t})},  \tilde{f}_\mathrm{N}\right)$ to exploit unaligned samples, and recall that $\tilde{f}_\mathrm{N}(\cdot)$ is the learned diffusion model. 

Our proposed vFL framework named \textsf{Diffu-AT} contains a federated branch ${f}(\cdot)$ and a local branch $\tilde{f}_\mathrm{L}(\cdot)$ for learning from aligned samples $\mathcal D_\mathrm{aligned}$ and unaligned samples $\tilde{\mathcal D}_\mathrm{unaligned}$ respectively in an alternative training fashion, and their bottom parts ${f}_\mathrm{L}^{(\mathrm{b})}$ and $\tilde{f}_\mathrm{L}^{(\mathrm{b})}$ share the parameter set. Specifically, at each training iteration, we randomly sample a mini-batch from  $\mathcal D_\mathrm{aligned}$ or $\tilde{\mathcal D}_\mathrm{unaligned}$, and then update the parameters of the corresponding branch. The sampling probability of selecting a mini-batch from $\mathcal D_\mathrm{aligned}$ is set to  $p= | {\mathcal D}_\mathrm{aligned} | / (| {\mathcal D}_\mathrm{aligned} |+|\tilde{\mathcal D}_\mathrm{unaligned}|)$. 

Put all together, Algorithm~\ref{alg} shows the training procedure of our \textsf{Diffu-AT}. Note that for large-scale deep learning based estimation models in online advertising and recommendation systems, the number of epochs is usually set to one.

\subsubsection{\textbf{Online Inference}} 
During online inference, only the federated branch $f(\cdot)$ is needed for producing the real-time predictions, and the local branch $\tilde{f}_\mathrm{L}(\cdot)$  is dropped. 

We notice that some studies focus on performing online inference based on a local model owned by label party via distilling knowledge from the federated model, motivated by reducing response time of receiving federated embedding from the non-label party~\cite{li2022jpl,li2022vflmpd}. However, in practice we found that the performance of the distilled local model drops drastically compared to the federated model, which is unacceptable in Alibaba's advertising business.

\begin{algorithm}[t]
\caption{Training Procedure of \textsf{Diffu-AT}}   
\label{alg}
\begin{algorithmic}[1]  
\REQUIRE \  1) Aligned samples $\mathcal D_\mathrm{aligned}=\{(\boldsymbol x_\mathrm{N}, \boldsymbol x_\mathrm{L}, \boldsymbol{y})\}$ \\ 
\qquad 2) Unaligned samples $\mathcal D_\mathrm{unaligned} = \left\{ \left(  \boldsymbol{x}_\mathrm{L}^{\mathrm{u}}, \boldsymbol{y}^{\mathrm{u}}  \right) \right\}$\\  
\ENSURE  The federated model $f(\cdot)$ composed of $\left({f}_\mathrm{L}^{(\mathrm{b})},  {f}_\mathrm{L}^{(\mathrm{t})},  {f}_\mathrm{N}\right)$ \\
\vspace{0.6em}
\STATE $\triangleright$ \textcolor{gray}{\textsf{Generating Federated Embeddings for Unaligned Samples}}
\STATE Learn a conditional diffusion model on  $\left\{ ({\boldsymbol{h}}_\mathrm{N}, {\boldsymbol{h}}_\mathrm{L}, \boldsymbol{y}) \right\}$
\STATE Inference on unaligned samples $\left\{\left(\boldsymbol{h}_\mathrm{L}^{\mathrm{u}}, {\boldsymbol{y}}^{\mathrm{u}}   \right)\right\}$, generating $\left\{ \tilde {\boldsymbol{h}}_\mathrm{N}^{\mathrm{u}} \right\}$
\STATE $\tilde{\mathcal D}_\mathrm{unaligned} = \left\{ \left( \tilde{\boldsymbol{h}}_\mathrm{N}^{\mathrm{u}}, \boldsymbol{x}_\mathrm{L}^{\mathrm{u}}, \boldsymbol{y}^{\mathrm{u}}  \right) \right\}$
\vspace{0.7em}
\STATE $\triangleright$ \textcolor{gray}{\textsf{Alternative Training}}
\STATE $p= | {\mathcal D}_\mathrm{aligned} | / (| {\mathcal D}_\mathrm{aligned} |+|\tilde{\mathcal D}_\mathrm{unaligned}|)$
\FORALL {epoch $e\gets 1$ \TO $E$}
  \FORALL {iteration $i\gets 1$ \TO $I$}  
    \STATE Randomly sampling a value $p_i$ from $\mathcal U(0, 1)$
    \IF {$p_i \leq p$}
        \STATE Yield a mini-batch from ${\mathcal D}_\mathrm{aligned}$
        \STATE Update the parameters of the federated branch $f(\cdot)$
    \ELSIF {$p_i > p$}
        \STATE Yield a mini-batch from $\tilde{\mathcal D}_\mathrm{unaligned}$
        \STATE Update the parameters of the local branch $\tilde{f}_\mathrm{L}(\cdot)$
    \ENDIF
  \ENDFOR 
\ENDFOR
\end{algorithmic}  
\end{algorithm}

\subsection{Defending Label Inference Attack}
As statement in Section~\ref{sec:vfl_issues}, due to the mathematical expression of the gradient $\boldsymbol{g}$ w.r.t. federated embedding contains label information $\boldsymbol{y}$, vFL models may suffer from potential label leakage, which means that a honest-but-curious non-label party can infer private labels. We focus on employ random perturbation methods to protect label information during vFL training.

Given the fact that the label leakage mainly comes from the difference between the {magnitudes and directions} of samples, an intuitive way to address such problem is to applying random convex combination of gradients while transmitting in the cut layer, also known as {mixup}~\cite{zhang2018mixup}. To alleviate label leakage, we propose a simple-yet-effective gradient mixup and projection approach named \textsf{MixPro},  which performs convex combinations and projections on in-batch sample gradients for protecting private label information. It employs mixup operation as initial perturbation to original gradients, and then performs projection to further remove useless information contained in original gradients. 

\textsf{MixPro} does not make any assumption about gradient distribution and can be seamlessly integrated into neural network based vFL framework. Next we first introduce the gradient mixup strategy, then adopt gradient projection to further modify the gradient directions to achieve better privacy-preserving performance during vFL model training. 

\subsubsection{\textbf{Gradient Mixup}}
For a batch of samples during training, we denote $\left\{\boldsymbol g_i\right\}_{i=1}^B$ to be the collection of gradients w.r.t. federated embeddings at the cut layer, where $B$ is the batch size. 
We formulate the mixup-based perturbed gradient of the $i$-th sample to be the convex combination of two sample gradients in the form below:
\begin{equation}
    {\boldsymbol g}_{mixed, i} = \lambda\cdot {\boldsymbol g}_{i} + (1-\lambda)\cdot {\boldsymbol g}_{r}
    \label{eq:perturb}
\end{equation}
where ${\boldsymbol g}_{r}$ is a random sample's gradient from the given batch. As stated in the original mixup method \cite{zhang2018mixup}, we choose $\lambda \sim \mathrm{Beta}(\alpha, \alpha)$ and set $\lambda > 0.5$, where $\alpha$ is a hyperparameter. 

Note that we set $\lambda > 0.5$ since the perturbed gradients should preserve more information from the original gradient including magnitude and direction to maintain the prediction performance of the vFL model. 
We only use the convex combination between two sample gradients in order to simplify the calculation, and we empirically found that this strategy also achieves similar performance with more gradients to be mixed.

\begin{figure}[t]
\vspace{1em}
\centering
\centerline{\includegraphics[width=0.7\columnwidth]{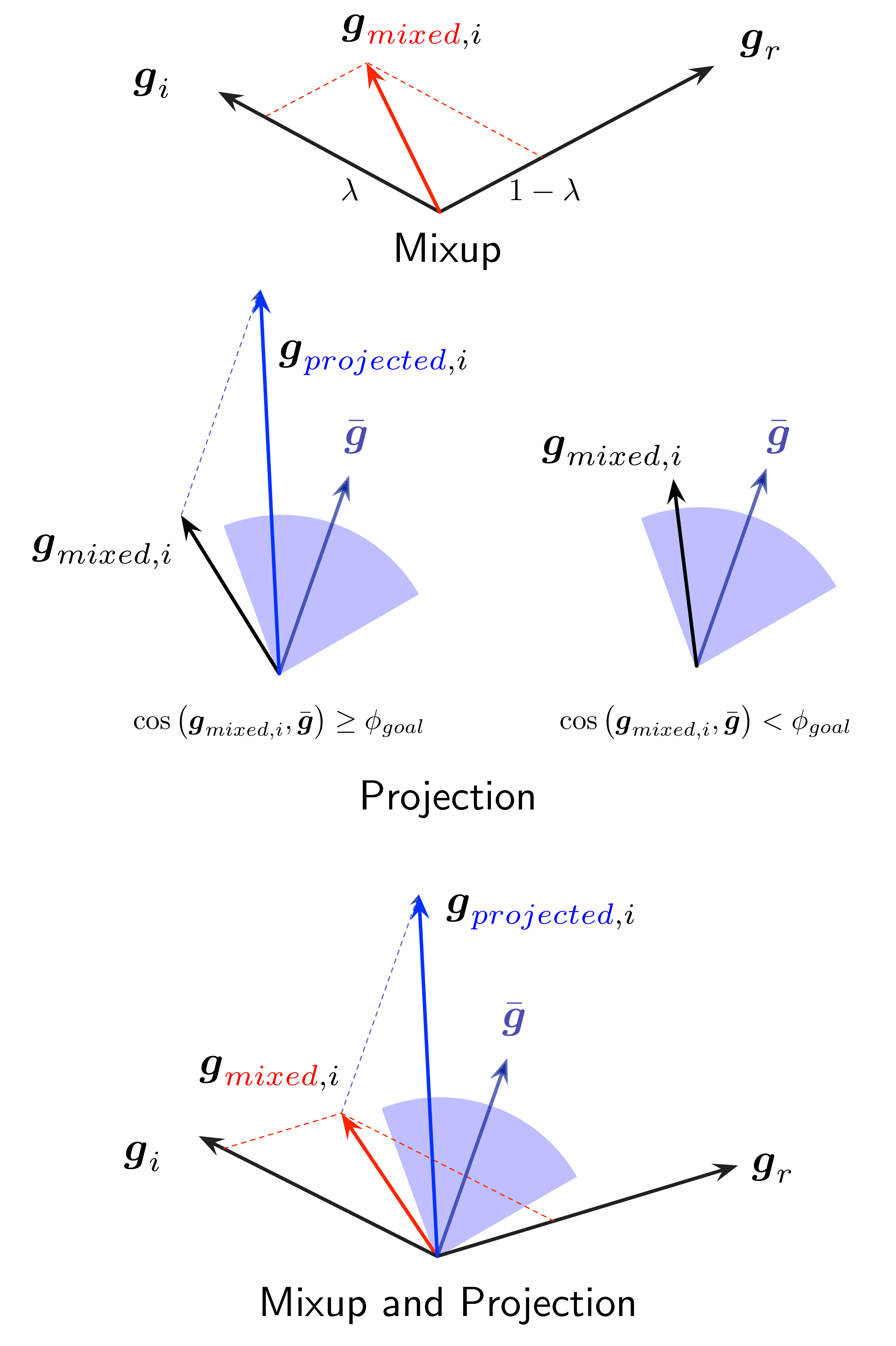}}
\caption{Illustration of gradient mixup and projection operations in \textsf{MixPro} for a training sample's gradient w.r.t. the federated embedding. }
\label{fig:mixpro}
\end{figure}

\subsubsection{\textbf{Gradient Projection}}\label{projection}
The above gradient mixup strategy may still suffer from the label leakage problem induced by the directions of gradients, since the direction information of the original gradient remains in the perturbed gradient to some extent. 

In order to keep the directions of gradients confined to a smaller region, we propose to further perform gradient projection on the mixed gradients. 
This is inspired by the studies in multi-task learning models~\cite{wang2021gradient, yu2020gradient}, in which the gradients for different tasks are projected along the direction of main one to avoid the conflict in gradient directions and thus achieve better performance. 
Given the intuition that higher similarity of gradients in orientation will effectively alleviate the label leakage from sample gradients, such projection technique in multi-task optimization can be further adopted in defending label leakage in vFL. 

Specifically, we denote $\Bar{\boldsymbol{g}} = \frac1B\sum_{i=1}^B \boldsymbol{g}_i$ as the average gradient for a batch of samples. 
We set the goal of the cosine similarity between the an in-batch gradient $\boldsymbol{g}_{mixed, i}$ and the average gradient $\Bar{\boldsymbol{g}}$ is that it should be larger than a pre-defined threshold $\phi_{goal}\in [-1, 1]$:
\begin{equation}
    \phi_{i} = \cos\left(\boldsymbol{g}_{mixed, i}, \Bar{\boldsymbol{g}}\right)  \geq \phi_{goal} \,.
\end{equation}

For a sample where the similarity goal is not achieved (that is, $\phi_{goal}>\phi_{i}$), the following projection operation will be applied and we obtain the projected gradient:
\begin{align}
    & \nonumber \boldsymbol{g}_{projected, i} = \boldsymbol{g}_{mixed, i} \  + \\
    & \frac{\lVert \boldsymbol{g}_{mixed, i} \rVert  \left(\phi_{goal}\sqrt{1-(\phi_{i})^2}-\phi_{i}\sqrt{1-(\phi_{goal})^2}\right)}{\lVert \Bar{\boldsymbol{g}} \rVert \sqrt{1-(\phi_{goal})^2}} \cdot \Bar{\boldsymbol{g}}.
\end{align}

And finally, the operation of our \textsf{MixPro} is:
\begin{align}
    \boldsymbol{g}_{perturbed, i} = 
    \begin{cases}
        \boldsymbol{g}_{mixed, i} \ , \  &\textrm{if}\ \phi_{i} \geq \phi_{goal} \ , \\
        \boldsymbol{g}_{projected, i} \ , \ &\textrm{if}\ \phi_{i} < \phi_{goal} \ .
    \end{cases}
\end{align}
During training process, based on our defense approach \textsf{MixPro}, for a batch of aligned training samples, the label party sends the perturbed gradients $\left\{\boldsymbol{g}_{perturbed,i}\right\}_{i=1}^B$ rather than the original ones $\left\{\boldsymbol{g}_i\right\}_{i=1}^B$ to the non-label party for updating the submodel $f_\mathrm{N}(\cdot)$, aiming to improve the privacy of the vFL model training. 

\textsf{MixPro} does not make any assumption about gradient distribution and can be seamlessly integrated into arbitrary neural network based vFL training  framework.

\section{Systematical Evaluations}\label{sec:exp}
As another contribution of our \textsf{FedAds} benchmark, we conduct systematical evaluations of various vFL models for both effectiveness and privacy aspects. 
For existing vFL algorithms that focus on improving effectiveness, we evaluate their performance in Section~\ref{exp:eff}. We then compare representative approaches for defending label inference attack, and  results are shown in Section~\ref{exp:privacy}.

\subsection{Experiments on Effectiveness}\label{exp:eff}
We first introduce experimental setup, evaluation metrics and comparative approaches in experiments, and then list implementation details and show experimental results. 

\subsubsection{\textbf{Experimental Setup.}}

We use our proposed dataset for evaluation. Specifically, we use 20\% of the training set as aligned samples $\mathcal{D}_\mathrm{aligned}$, and the rest 80\% of the training set is used as unaligned samples $\mathcal{D}_\mathrm{unaligned}$ by removing their features from non-label party. The performance is evaluated using the test set. 

This setup allows us to know the upper bound of exploiting unaligned samples in vFL on our dataset: if we use the full training data as aligned samples and train a vFL model, the model's performance is the upper bound because we ``leak'' the non-label party's features of $\mathcal{D}_\mathrm{unaligned}$.

\subsubsection{\textbf{Evaluation Metrics.}}
We use AUC and negative log likelihood (NLL for short) as the evaluation metrics for effectiveness. The former  measures ranking performance on candidates, and the latter reflects calibration performance of predicted scores.

\subsubsection{\textbf{Comparative Approaches.}}
We compare the following approaches to evaluate effectiveness:
\begin{itemize}
    \item \textsf{Local}\quad is a model trained on label party's features only, without using any features from non-label party. 
    \item \textsf{VanillaVFL}\quad is a vFL model trained on aligned samples using the loss in Equation~\ref{eq:vfl_loss}. 
    \item \textsf{HeuristicVFL}\quad further exploits label party's unaligned samples, in which the missing non-label party's features are synthesized with a heuristic way: for each unaligned sample, we retrieve the user ID from aligned samples, and compute the averaged federated embedding of this user in \textsf{VanillaVFL}. Then we perform alternative training (see Section~\ref{sec:at}).
    \item \textsf{SS-VFL}~\cite{castiglia2022SSVFL}\quad exploits unaligned samples with self-supervised learning. Each party first employs its local samples to perform unsupervised pretraining, and then a \textsf{VanillaVFL} is trained as the final model.
    \item \textsf{FedCVT}~\cite{kang2022fedcvt}\quad exploits unaligned samples with both self-supervised learning and semi-supervised learning. It first performs unsupervised pretraining to learn a two-stream network. Then a similarity function is used to generate unaligned samples' features. Finally it combines unlabeled unaligned samples and labeled aligned samples with semi-supervised learning in a co-training fashion.\footnote{Note that the time complexity of similarity computing process in \textsf{FedCVT} is too high, and thus we re-implement it with an approximate way through randomly dropping some candidates before similarity computing.}
    \item \textsf{VFL-MPD}~\cite{li2022vflmpd}\quad exploits unaligned samples with a specific self-supervised learning task, where a matched pair detection task is proposed to learn powerful representations using large size of unaligned samples. 
    \item \textsf{FedHSSL}~\cite{he2022FedHSSL}\quad exploits unaligned samples with two-stage pretraining. The first stage is cross-party pretraining that fits learned representations each other. The second stage is local pretraining where the learning objective is based on data augmentation. 
    \item \textsf{JPL}~\cite{li2022jpl}\quad exploits unaligned samples via synthesizing non-label party's features based on learning a mapping function between label party's features and non-label party's features, with several constraints like representation equivalence and label  discrimination.\footnote{Note that the original \textsf{JPL} approach distills the knowledge of federated model to a local model, which results in performance drop. For fair comparison, we replace the distillation with the alternative training as we stated in Section~\ref{sec:at}.}
    \item \textsf{Diffu-AT}\quad is our proposed approach. 
    \item \textsf{ORALE}\quad is a model trained on the full training set that the non-label party's features of unaligned samples are known in advance. Therefore its performance is the upper bound of exploiting unaligned samples in our dataset. 
\end{itemize}

\begin{table}[t]
    \centering
    \caption{Effectiveness evaluation of comparative vFL algorithms for CVR estimation.}
    \begin{tabular}{lccccc}
        \toprule 
        {\textbf{Method}} & {\textbf{Ranking}: AUC} & {\textbf{Calibration}: NLL} \\ 
        \midrule
        \multicolumn{5}{l}{\textit{Label party's features} $\left\{(\boldsymbol{x}_\mathrm{L}, \boldsymbol{y})\right\}$ } \\
        \quad\textsf{Local}  & 0.609  & 0.0391 \\
        \cmidrule(lr){1-5}
        \multicolumn{5}{l}{\textit{+ vFL training} $\left\{(\boldsymbol{x}_\mathrm{N},\boldsymbol{x}_\mathrm{L}, \boldsymbol{y})\right\}$}\\
        \quad\textsf{VanillaVFL}    & 0.620  & 0.0389 \\
        \cmidrule(lr){1-5}
        \multicolumn{5}{l}{\textit{+ Unaligned samples} $\left\{ \left( \boldsymbol{x}_\mathrm{L}^{\mathrm{u}}, \boldsymbol{y}^{\mathrm{u}}  \right) \right\}$ }\\
        \quad\textsf{HeuristicVFL} & 0.630    &  0.0387\\
        \quad\textsf{SS-VFL} & 0.636    &  0.0381\\  
        \quad\textsf{FedCVT}  & 0.639    &  0.0379\\               
        \quad\textsf{VFL-MPD}  & 0.641    &  \textbf{0.0373}\\  
        \quad\textsf{FedHSSL} & 0.642    &  0.0375\\  
        \quad\textsf{JPL}    & 0.644    &  0.0374 \\
        \quad\textsf{Diffu-AT} (Ours)   & \textbf{0.645}    &  0.0375 \\    
        \midrule
        \multicolumn{5}{l}{\textit{Upper bound} $\left\{ \left( \boldsymbol{x}_\mathrm{N}^{\mathrm{u}}, \boldsymbol{x}_\mathrm{L}^{\mathrm{u}}, \boldsymbol{y}^{\mathrm{u}}  \right) \right\}$ }\\
        \quad\textsf{ORALE}  & 0.658 & 0.0367 \\
        \bottomrule
    \end{tabular}
    \label{exp:effectiveness}
\end{table}

\subsubsection{\textbf{Implementation Details.}}
We choose YouTube-DNN~\cite{covington2016deep} as the backbone. Each feature is represented as an 8-dim embedding. The non-label party submodel $f_\mathrm{N}$ contains an embedding table and a two-layered DNN with output sizes of [128, 32]. The label party's submodel contains two branches $f_\mathrm{L}$ and $\tilde{f}_\mathrm{L}$, where they share the embedding table and bottom part (a two-layered DNN with output sizes of [256, 128]), and each top part is a single layer with logistic function to produce predicted score. The two branches do not share the batch normalization operations. The batch size is set to 256. In the conditional diffusion model, the total step $T$ is set to 1000 as in previous work~\cite{ho2020denoising}. For timestep $t$ we use sine and cosine functions to encode, and the schedule for $\beta$ is linear schedule. We implement the models using \textsf{XDL}\footnote{\url{https://github.com/alibaba/x-deeplearning}} and \textsf{EFLS}.

\subsubsection{\textbf{Experimental Results.}}
Table~\ref{exp:effectiveness} shows the evaluation results for effectiveness of all the comparative approaches. Note that 0.3\% improvement on AUC can be regarded as a large improvement in large-scale industrial datasets~\cite{cheng2016wide}. We observe that \textsf{VanillaVFL} outperforms \textsf{Local} by around 1.1\% AUC, which demonstrates that incorporating non-label party's features to perform federated training is effective to improve the estimation performance. 
Furthermore, exploiting unaligned samples usually boost the AUC and NLL compared to \textsf{VanillaVFL}. For instance, the simple approach \textsf{HeuristicVFL} outperforms \textsf{VanillaVFL} by 1.0\% AUC which shows the potential of this direction. 
By comparing the approaches that exploits unaligned samples, we can see that the models considering label information of unaligned data (such as \textsf{Diffu-AT} and \textsf{JPL}) generally perform better than the self-supervised learning models (such as \textsf{FedHSSL}, \textsf{VFL-MPD} and \textsf{SS-VFL}), and thus the labeled samples from label party are the key for improving traditional vFL. 

Our proposed \textsf{Diffu-AT} shows the best performance on the ranking ability among all comparative approaches, verifying that the synthesized federated embeddings with diffusion model can enhance the representation of unaligned samples. We also see that \textsf{VFL-MPD} performs the best on the calibration performance, and we suggest that its pretraining objective obtains better model initialization. We leave the improvement of \textsf{Diffu-AT}'s  calibration performance in future work.

\subsection{Experiments on Privacy}\label{exp:privacy}

\subsubsection{\textbf{Evaluation Metrics for Privacy.}}
We first introduce how a non-label party can perform label inference attack to steal private label information, and then gives the evaluation metrics for privacy. 

\vspace{0.5em}
\textbf{Label inference attack.}

Specifically, the non-label party is the \textbf{attacker} that performs label inference attack.
The objective of an attacker is to infer the private labels ${\boldsymbol{y}}$ owned by the label party based on exchanged federated embeddings $\boldsymbol{h}_\mathrm{N}$ and/or gradients $\boldsymbol{g}$. Specifically, the attack can be performed between two iterations during training phrase, or after the training is finished. 
We assume that the non-label party is honest-but-curious, which means that it \textit{cannot} interfere with the training process (such as sending wrong hidden representations to the label party). Under this situation, the non-label party can employ arbitrary classifiers to infer labels.

From Equation~\ref{eq:gradient} we observe that the form of gradient $\boldsymbol{g}$ actually contains label information ${\boldsymbol{y}}$. Besides,~\citet{sun2022label} also found that the federated embeddings gradually have correlations with labels during vFL training. We introduce two attack strategies based on gradient and federated embedding respectively. 
\begin{enumerate}
    \item Gradient-based attack~\cite{li2022label}. Given the observation that a model tends to be less confident about ``a positive sample being positive'' than ``a negative sample being negative'', if a sample's gradient norm $\lVert \boldsymbol{g} \rVert_2$ is larger than a threshold, the attacker infers that it belongs to positive class. 
    \item Federated embedding-based attack~\cite{sun2022label}. After vFL model training, we perform clustering on the federated embeddings to place all training samples into two clusters. Given the prior knowledge that positive samples' size is smaller than that of negative samples, the attacker infers that the samples in the smaller cluster belong to the positive class. 
\end{enumerate}

\vspace{0.5em}
\textbf{Evaluation metrics.}

For a vFL model equipping defense approach, we evaluate two aspects including \textit{utility} and \textit{privacy}. 
The utility aspect is about the prediction performance of the model, and we choose AUC used in previous experiments on effectiveness. 

The privacy aspect evaluates the defense ability to label inference attack. Because the evaluation relies on the attack strategy that may be stochastic to some extent, we design a metric named $\Delta\mathrm{LeakAUC}$ to compute relative improvement to a base model: 1) we first use the model without any defense approach as the base, and compute the AUC value given the labels inferred by attacker and the true labels, namely $\mathrm{LeakAUC}_\mathrm{base}$. 2) Consider a model with a specific defense approach, similarly we compute the $\mathrm{LeakAUC}_\mathrm{exp}$. 3) Finally we compute the relative improvement $\Delta\mathrm{LeakAUC} = (\mathrm{LeakAUC}_\mathrm{exp} - \mathrm{LeakAUC}_\mathrm{base}) / \mathrm{LeakAUC}_\mathrm{base} $ as the evaluation metric. 
For the LeakAUC metric, \textbf{lower} is better, because it is expected that the attacker cannot recover true labels. 

\subsubsection{\textbf{Comparative Approaches.}}

We compare the following random perturbation based defense approaches. 
\begin{itemize}
    \item \textsf{No\ Defense}\quad is the vFL model that does not equip any defense approach during training. 
    \item \textsf{DP}~\cite{abadi2016deep}\quad employs differential privacy to obtain a generic gradient perturbation framework, in which DP is enforced on the transmitted information. 
    \item \textsf{Marvell}~\cite{li2022label}\quad adds Gaussian noise to perturb original gradients, where the covariance matrices of noise distribution are computed based on minimizing leakage level under a noise power constraint. The noise form assumes that the original gradient also follows Gaussian distribution. 
    \item \textsf{MixPro}\quad is our proposed defense approach. 
\end{itemize}

\begin{table}[t]
    \centering
    \caption{Privacy evaluation of comparative defense approaches to label inference attack.}
    \begin{tabular}{lccccc}
        \toprule 
         {\textbf{Method}} & {\textbf{Privacy}: $\Delta\mathrm{LeakAUC}$} & {\textbf{Utility}: AUC} \\ 
        \midrule
        \textsf{No\ Defense} & - &\textbf{0.620}\\
        \textsf{DP} &-5.7\%&0.602\\
        \textsf{Marvell} &\textbf{-24.7\%}&0.601\\
        \cmidrule{1-3}
        \textsf{MixPro} (Ours)                &-11.3\%&0.602   \\
        \bottomrule
    \end{tabular}
    \label{tab:privacy}
\end{table}

\subsubsection{\textbf{Implementation Details.}}
We choose the \textsf{VanillaVFL} model used in previous experiments on effectiveness as the base model to compute $\mathrm{LeakAUC}_\mathrm{base}$, and the label inference attack strategy is the federated embedding-based attacker. 
There are two key hyperparameters in our proposed \textsf{MixPro}: $\alpha$ to control the gradient mixup strategy and $\phi_{goal}$ to determine how the mixed gradient should be projected. Generally, a larger $\alpha$ will force the mixup strategy to show less uncertainty and the mixup weight $\lambda$ will thus become closer to 0.5, leading to better privacy preservation but greater compromise on AUC. A larger $\phi_{goal}$ will also narrow down the region of gradients' directions and provide better privacy performance with more trade-off on AUC. In experiments we set $\alpha=0.6$ and $\phi_{goal}=\sqrt{3}/2$ by default.

\subsubsection{\textbf{Experimental Results.}}
Table~\ref{tab:privacy} shows the experimental results on our proposed datasets of our proposed \textsf{MixPro} and other compared approaches against the label inference attack. The straightforward \textsf{DP} only achieves very limited privacy performance compared to \textsf{No\ Defense}, which means that more sophisticated approaches are needed for protecting label information. 

\textsf{Marvell} reduces the LeakAUC by a large margin and achieves an acceptable level, demonstrating that the a well-designed random perturbation strategy is very effective for defending label inference attack. 
Our proposed \textsf{MixPro} performs much better than \textsf{DP}, verifying its privacy performance in vFL model training, and \textsf{Marvell} is still the state-of-the-art defense approach. We suggest that in \textsf{MixPro} the random sampling operation for mixup and projection may restrict its defense ability, because the size of positive samples is very small and thus two combined gradients usually come from the same class. 
We also notice that both \textsf{MixPro} and \textsf{Marvell} result in a drop of around 2.0\% AUC compared to \textsf{No\ Defense}. Therefore, better utility-privacy trade-off is a key direction in further defense approaches during vFL model training.

\section{Related Work}\label{related work}
There are many great benchmarks for FL, such as LEAF~\cite{caldas2018leaf}, FedML~\cite{he2020fedml}, Flower~\cite{beutel2020flower}, FedScale~\cite{lai2022fedscale}, FLamby~\cite{terrail2022flamby} and pFL-bench~\cite{chen2022pfl}. 
They mainly focus on horizontal FL and personalized FL, and to our knowledge no vFL benchmarks have been proposed for fair comparing existing approaches, especially for neural network based vFL approaches.
In this work, we propose the vFL benchmark named \textsf{FedAds}, which provides a large-scale dataset collected from Alibaba's advertising system, as well as systematical evaluations for existing approaches. Therefore we believe that the \textsf{FedAds} makes good contributions to facilitate vFL research.

\section{Conclusion and Future Work}
We introduce \textsf{FedAds}, the first benchmark for privacy-preserving CVR estimation, to facilitate systematical evaluations for vFL algorithms. It contains 1) a large-scale real-world dataset from our online advertising platform, collected from an ad delivery business relying on vFL-based ranking models, as well as 2) systematical evaluations for both effectiveness and privacy aspects of various neural network based vFL algorithms through extensive experiments. Besides, to improve vFL effectiveness, we explore to incorporate unaligned data via generating unaligned samples' feature representations using generative models. To protect privacy well, we also develop perturbation based on mixup and projection operations. Experiments show that they achieve reasonable performances. 

In future work, we shall explore the following directions: 1) Improving the calibration performance of vFL models~\cite{pan2020field,wei2022posterior}. 2) Alleviating the sample selection bias issue in CVR estimation models through debiasing approaches~\cite{guo2021enhanced,xu2022ukd} for vFL models. 3) Improving vFL training efficiency. 4) Extending the usage of vFL fashion from ranking stage to retrieval stage in online adverting systems.

\section*{Acknowledgement}
This work was supported by Alibaba Innovative Research project.  We thank all the anonymous reviewers for their valuable comments.
We also thank Jinquan Liu and Prof. Baoyuan Wu for assistance. 

\balance
\bibliographystyle{ACM-Reference-Format}
\bibliography{src-shortversion.bib}

\end{document}